\documentclass[sigconf]{acmart}
 \UseRawInputEncoding

\AtBeginDocument{}

\copyrightyear{2024}
\acmYear{2024}
\setcopyright{rightsretained}
\acmConference[ICPC '24]{32nd IEEE/ACM International Conference on Program Comprehension}{April 15--16, 2024}{Lisbon, Portugal}
\acmBooktitle{32nd IEEE/ACM International Conference on Program Comprehension (ICPC '24), April 15--16, 2024, Lisbon, Portugal}\acmDOI{10.1145/3643916.3644409}
\acmISBN{979-8-4007-0586-1/24/04}

\usepackage{enumitem}
\usepackage{multirow}
\usepackage{xspace}
\usepackage{mdframed}
\usepackage{color}
\usepackage{colortbl}
\usepackage{fancyvrb}
\usepackage{url}
\usepackage{booktabs}

\begin{document}

\title{Analyzing Prompt Influence on Automated Method Generation: An Empirical Study with Copilot}

\author{Ionut Daniel Fagadau}
\affiliation{\institution{University of Milano - Bicocca}
  \city{Milan}
  \country{Italy}
}
\email{i.fagadau@campus.unimib.it}

\author{Leonardo Mariani}
\affiliation{\institution{University of Milano - Bicocca}
  \city{Milan}
  \country{Italy}
}
\email{leonardo.mariani@unimib.it}

\author{Daniela Micucci}
\affiliation{\institution{University of Milano - Bicocca}
  \city{Milan}
  \country{Italy}
}
\email{daniela.micucci@unimib.it}

\author{Oliviero Riganelli}
\affiliation{\institution{University of Milano - Bicocca}
  \city{Milan}
  \country{Italy}
}
\email{oliviero.riganelli@unimib.it}

\newcommand{\LEO}[1]{\textcolor{blue}{{\it [Leonardo says: #1]}}}
\newcommand{\DAN}[1]{\textcolor{red}{{\it [Daniela says: #1]}}}

\newenvironment{todo}{\color{red}}{\color{black}}

\newcommand{\numFeatures}{eight\xspace}
\newcommand{\numProblems}{\textcolor{red}{TBD}\xspace}
\newcommand{\totalQueries}{$124,800$\xspace}

\newcommand{\prompt}[1]{{\small \textit{#1}\xspace}}
\newcommand{\urlRepo}{\url{https://shorturl.at/hmpBM}\xspace}

\begin{abstract}
Generative AI is changing the way developers interact with software systems, providing services that can produce and deliver new content, crafted to satisfy the actual needs of developers. For instance, developers can ask for new code directly from within their IDEs by writing natural language prompts, and integrated services based on generative AI, such as Copilot, immediately respond to prompts by providing ready-to-use code snippets. Formulating the prompt appropriately, and incorporating the useful information while avoiding any information overload, can be an important factor in obtaining the right piece of code. The task of designing good prompts is known as prompt engineering.

In this paper, we systematically investigate the influence of eight prompt features on the style and the content of prompts, on the level of correctness, complexity, size, and similarity to the developers' code of the generated code. We specifically consider the task of using Copilot with 124,800 prompts obtained by systematically combining the eight considered prompt features to generate the implementation of 200 Java methods. 
Results show how some prompt features, such as the presence of examples and the summary of the purpose of the method, can significantly influence the quality of the result.
\end{abstract}



\maketitle

\section{Introduction} \label{sec:introduction}

Generative AI solutions, and in particular those built on Large Language Models (LLM), such as ChatGPT~\cite{ChatGPT2023}, Bard~\cite{GoogleBard2023}, and CoPilot~\cite{Copilot2023}, promise to become a powerful tool that can aid software developers in completing their tasks more efficiently and effectively. For instance, LLMs have already been exploited to support programming tasks by automatically generating code that responds to a given natural language request formulated by a user~\cite{Li:Alphacode:Science:2022,Nijkamp:CodeGen:ICLR:2023,Fried:InCoder:ICLR:2023}. Some other recent studies investigated how developers interact with these tools during project development~\cite{Barke:CoPilotGroundTheory:ACMPL:2023}, and how usable these tools are~\cite{Barke:CoPilotGroundTheory:ACMPL:2023}. 

A key concern about LLMs is that they are known to be sensitive to the prompt, that is, the quality of the result strongly depends on the query that is asked by the user to the model~\cite{Rodriguez:Prompts:REW:2023,Liu:PromptImages:CHI:2022,Lo:ArtPrompting:IRSQ:2023}.  In fact, the style of the prompt, as well as its content, may determine the level of correctness of the response. This is also confirmed by our observations in the domain of code generation. 
For instance, when asking for the body of the Java method \texttt{String convertToBase7(int num)} Copilot generates better code when the prompt includes some basic hints about the semantics of the intended solution, such as  "\prompt{The base 7 digits are 0‒6 and the digit positions represent powers of 7}". Or sometimes the resulting code might be better by just considering some simple stylistic rules, such as using the imperative mood (e.g., \prompt{\ldots implement pattern matching that...}), which is often easier to be interpreted by LLMs compared to future (e.g., \prompt{\ldots the method shall implement pattern matching that...}).
As a consequence, researchers are actively studying how to interact with LLMs, so that the prompts that are likely to provide the best results can be quickly formulated and submitted to the models, without losing time with ineffective interactions. 

Only a few studies preliminary investigated prompt engineering for code generation. Denny et al. studied the impact of changes implemented in prompts used for solving Python programming problems~\cite{Denny:ConversingCoPilot:TSCSE:2023}. The results show that prompt engineering has been useful in the vast majority of the cases to improve the correctness of the resulting code. White et al. proposed patterns that can be used by programmers to write prompts in different situations, however without reporting data about their effectiveness~\cite{White:PromptPatterns:arXiv:2023}. Ren et al. investigated the specific case of adjusting the prompt to obtain proper exception handling code~\cite{Ren:CodeGen:ASE:2023}. So far, no study systematically considered prompt engineering in code generation tasks.

In this paper, we propose a first controlled experiment that investigates the effectiveness of \numFeatures prompt features on the same prompt for $200$ Java code generation problems extracted from both GitHub~\cite{GitHub2023} and LeetCode~\cite{LeetCode2023}. For example, we systematically compare the generated code when the same request is formulated according to different grammar styles (e.g., active and passive forms) or includes different content (e.g., with and without examples). We consider both GitHub and LeetCode to investigate the impact of the same prompt features when occurring within different kinds of prompts. Prompts derived from the GitHub methods' comments usually are concise explanations of a method's behavior, such as \prompt{ The function inserts the given vertex into a sorted position in the given array}. On the contrary, prompts derived from LeetCode methods' descriptions usually are lengthy descriptions about the method to be implemented, such as 
\prompt{The function implements wildcard pattern matching \ldots where '?' matches a single character \ldots return true if any match is found. Example: \ldots}.

In our study, we focus on the Copilot-3 LLM, since it is the most adopted AI developer tool, with more than 1 million developers and 20,000 organizations that adopted it~\cite{CoPilotAdoption}.  
We executed Copilot with a total of \totalQueries queries discovering important information about prompt engineering for code generation. In particular, we discover that including a summary of the purpose of the method and including examples in the prompt are particularly useful to obtain code that passes the available test cases. Some stylistic rules may also have an impact. For instance, the usage of the presence tense seems beneficial in prompts. 

On the other hand, including some other information has not produced beneficial results. For instance, the inclusion of boundary cases and contextual information in the prompts had no significant effect on the level of correctness of the results. Including excessive information may even have a negative effect, as reported for boundary cases that made Copilot produce code that differs more from the code implemented by the developers when included in the prompt.  

The experimental materials necessary to reproduce our study have been made available for online access at \urlRepo.

The paper is structured as follows. Section~\ref{sec:prompt} introduces prompt engineering and describes the prompt features that we considered in our experimental study. Section~\ref{sec:methodology} reports the three research questions that we investigated and describes the methodology that we followed to answer them. Section~\ref{sec:results} presents the results that we obtained for each research question, discusses threats to validity, and distills some advice about the definition of the prompts. Section~\ref{sec:related} discusses related work. Section~\ref{sec:conclusions} provides final remarks.  \section{Prompt Engineering and Prompt Features} \label{sec:prompt}

Although powerful, LLMs may generate results of different quality depending on the style and content of the prompt. For this reason prompt engineering, that is, ``the formal search for prompts that retrieve
desired outcomes from language models''~\cite{Liu:PromptImages:CHI:2022}, is now an active field of research. For example, in computer vision, studies on prompt engineering revealed that focusing on the mood and style of the keywords that occur in the prompts is more important than rephrasing the prompts themselves~\cite{Liu:PromptImages:CHI:2022}. 

In a nutshell, prompt engineering can be seen as a kind of programming in natural language, where programming statements are the natural language sentences in the prompts~\cite{Reynolds:PromptProrammin:CHI:2021}. Understanding how to write good prompts is indeed also relevant when considering code generation tasks. To derive insights about how to write prompts for code generation, we systematically considered the same prompts written according to different styles and with variations in the content. The selection of the features derives from the analysis of the comments and prompts present in the GitHub and LeetCode methods we extracted. In particular, we considered three prompt features affecting the \emph{style} of the sentences and five prompt features affecting the \emph{content} of the prompt, for a total of  \numFeatures prompt features.

\smallskip
\noindent \textbf{Prompt features about the \emph{style}}

\noindent \textbf{Mood}:  \textit{(a) Indicative}, when the prompt uses the indicative mood to specify what the function has to do, for instance \prompt{Given an integer array \ldots, the function returns true if there are two distinct indices \ldots}; \textit{(b) Imperative}, when the prompt uses the imperative mood of to specify what the function has to do, for instance \prompt{Given an integer array \ldots, return true if there are two distinct indices \ldots};

\noindent \textbf{Sentence Mode}: \textit{(a) Active}, when the prompt is in active mode, for instance \prompt{Given an integer array \ldots, the function returns True if \ldots}; \textit{(b)~Passive}, when the prompt is in passive mode, for instance \prompt{An integer array and \ldots are given to the function. True is returned by the function if \ldots}.

\noindent \textbf{Tense}: \textit{(a) Present}, when the prompt uses the present tense, for instance \prompt{Given an integer array \ldots, the function returns true if there are two distinct indices \ldots}; \textit{(b) Future}, when the prompt uses the future tense, for instance \prompt{Given an integer array \ldots, the function will return true if there are two distinct indices \ldots}

\smallskip
\noindent \textbf{Prompt features about the \emph{content}}

\noindent  \textbf{Reference to the Parameters}: \textit{(a) Implicit without names}, when the prompt implicitly refers to the parameters of the functions without using their name, for instance \prompt{Return True if there are two distinct indices \ldots in an integer array, such that \ldots};  \textit{(b) Implicit with names}, when the prompt implicitly refers to the parameters but uses their names, for instance \prompt{Return True if there are two distinct indices \ldots in an integer array nums, such that \ldots}; \textit{(c) Explicit without names}: when the prompt explicitly refers to the parameters without using their names, for instance \prompt{Given an integer array and \ldots, return True if there are two distinct \ldots};  \textit{(d) Explicit with names}: when the prompt explicitly refers to the parameters using their names, for instance \prompt{Given an integer array nums and \ldots, return True if there are two distinct \ldots}.

\noindent  \textbf{Boundary Cases}: \textit{(a) Missing}, when the prompt does not include any boundary case; \textit{(b) Implicit}, when the boundary case is stated but it is not stated what the method should do, for instance \prompt{Given an integer array \ldots return true if there are \ldots and the array is not null}; \textit{(c) Explicit}, when the boundary case is stated, including the expected behavior, for instance \prompt{Given an integer array \ldots return true if there are \ldots. If the array is null, return false}.

\noindent  \textbf{Summary of the Method}: \textit{(a) Missing}, when the semantic of the method is not summarized in the prompt, like in all the examples reported so far; \textit{(b) Provided Upfront}, when the semantic is stated at the beginning of the prompt, for instance \prompt{The result is a boolean representing if there are any duplicates in the array. Given an integer \ldots}; \textit{(c) Provided Afterward}, when the semantic of the method is summarized at the end of the prompt, for instance \prompt{Given an integer \ldots The result is a boolean representing if there are any duplicates in the array}.

\noindent  \textbf{Examples}: \textit{(a) Present}, when the prompt includes examples about the input-output behavior of the method, for instance \prompt{Given an integer array \ldots Example Input nums=[1, 2, 3, 1], Output = true}; \textit{(b) Absent}, when the prompt does not include any example.

\noindent \textbf{Context}:  \textit{(a) Present}, when the prompt includes information about the characteristics of the inputs and the outputs, for instance \prompt{Given an integer array \ldots Constraints 1 <= nums.length <= 100000, 0<=k<=100000}; \textit{(b) Absent}, when the prompt does not include such information. \section{Methodology} \label{sec:methodology}

The objective of our study is to investigate how prompt features may impact the correctness of the code generated by LLMs. In particular, we consider the generation of a method body as a task, and Copilot as LLM assisting this task. We selected this task because these tools are mainly designed and trained to help developers with the implementation of code snippets, such as individual methods and functions~\cite{Barke:CoPilotGroundTheory:ACMPL:2023,CopilotUsage}. We selected Copilot because it is the most popular and used AI-based code assistant at the moment~\cite{CoPilotAdoption}. 

Our study is structured around three main research questions:

\noindent \textbf{RQ1 - How do prompt features impact on the \emph{correctness} of the code generated by Copilot?} This research question investigates how prompt features impact on the generation of code that compiles and passes test cases.

\noindent \textbf{RQ2 - How do prompt features impact on the \emph{complexity} of the code generated by Copilot?} This research question investigates how prompt features impact on the cyclomatic complexity and size of the generated code.  

\noindent \textbf{RQ3 - How do prompt features impact on the \emph{similarity} between the code generated by Copilot and the code implemented by the developers?} This research question investigates how prompt features impact the possibility of obtaining code that is syntactically and semantically close to the code that the developers have implemented.

We describe below how we selected the prompts for answering RQ1-3, how we obtained the alternative prompts, consistently with the identified prompt features, how we collected the responses from Copilot, and finally how we analyzed the results. 

\subsection{Selection of the Prompts}
Our experiment involves creating many different variants for the same prompts. We have thus to limit the initial set of prompts to start from to achieve a feasible total number of prompts to be analyzed. On the other hand, we want to study prompt engineering both in the context of short prompts mostly reflecting what developers already write as comments for their code, and in the context of richer prompts designed to be more explicative of the code that must be generated. For this reason, we decided to select 100 GitHub methods with comments, representing the case of shorter prompts derived from comments, and 100 LeetCode\footnote{LeetCode is an online platform for programming challenges, and to train developers to get ready for technical job interviews.} methods, representing the case of longer and more explicative prompts.    

To select the prompts from GitHub, we built on top of results obtained by Mastropaolo et al.~\cite{Mastropaolo:RobustnessGenCode:ICSE:2023} who collected prompts from the Javadoc comments present in the GitHub code that is part of Java projects with at least 300 commits, 50 contributors, and 25 stars. We selected prompts corresponding to methods with at least 75\% of the statements exercised by the test case available in the repository of the project. From this set of prompts and methods, to make sure to consider a variety of prompts of different lengths, we randomly selected 40 prompts in the range 36-97 characters, 30 prompts in the range 98-159 characters, and 30 prompts in the range 160-221 characters\footnote{36 characters are the length of the shortest prompt in the dataset and 221 characters are the length of the longest prompt.}.

To select the prompts from LeetCode, we exploited the classification of the problems present in the platform as easy, medium, and hard tasks. We thus randomly selected 40 easy problems, 30 medium problems, and 30 hard problems. Differently from GitHub, LeetCode provides the prompts but provides neither the expected solution nor the test suite to validate the code\footnote{LeetCode provides test suites to validate solutions, but the test suites are not publicly available and they can be used only by submitting solutions online, which is not doable for such a large study.}. To obtain the solutions for the selected problems, we exploited the walkccc platform ({\small \url{https://walkccc.me/LeetCode}}) that stores solutions to LeetCode problems. Finally, to obtain the test cases, we automatically generated the tests from the solutions by running the EvoSuite test generator tool ({\small \url{https://www.evosuite.org}}) with its default configuration. We obtained a total of 426 test cases, an average number of 4.3 test cases per method. The average statement coverage achieved by these tests is 99.8\%. Actually 97 methods are fully covered, and three methods have a lower coverage, with a minimum equal to 88\%.

At the end of this step, we obtained 100 prompts derived from GitHub and 100 prompts derived from LeetCode. Each prompt is associated with a method signature, a method implementation, and a set of test cases. In all cases, we packaged the code as a Maven project to ease the automation of the next steps.

\subsection{Generation of the Alternative Prompts}
So far, the collected prompts are available in a single shape in terms of prompt features, while our objective is to systematically study the impact of different combinations of prompt features. To this end, we created every possible version of each prompt, according to the prompt features reported in Section~\ref{sec:prompt}. 

In the case of the prompts extracted from GitHub, five prompt features can be modified systematically: the three prompt features about \emph{style}, plus the \emph{Reference to the Parameters} and the \emph{Boundary Cases}. The prompt features \emph{Summary of the Method}, \emph{Examples}, and \emph{Context} are always absent, and we decided to not add this information ourselves in the prompts to avoid any bias in the experiment. We, however, modified the prompts considering any possible combination of the applicable features obtaining 96 alternative versions of each prompt (2 \emph{Mood} $\times$ 2 \emph{Sentence Mode} $\times$ 2 \emph{Tense} $\times$ 4 \emph{Reference to the Parameters} $\times$ 3 \emph{Boundary Cases}), for a total of $9,600$ prompts studied.

In the case of the prompts extracted from LeetCode, all prompt features can be modified systematically. This leads to 1,152 prompt variants (96 prompt variants as for GitHub prompts $\times$ 3 \emph{Summary of the Method} $\times$ 2 \emph{Examples} $\times$ 2 \emph{Context}) for each prompt, for a total of 115,200 prompts studied. 

We obtained these so many variants of the prompts with a combination of manual effort and automatic scripts. Changes to prompt features like \emph{Mood}, \emph{Sentence Mode}, and \emph{Reference to the Parameters}, require manual intervention, since part of the sentence should be rewritten. These changes have been actuated first. The remaining prompt features could be addressed semi-automatically, that is, for each prompt that has to be modified, we implemented an ad-hoc script that makes substitutions and suppressions, finally obtaining the modified prompts. 

Note that we always generated the variants of the features manually, and the scripts have been used to only systematically combine their values. For instance, the scripts can include the right instance of a feature in a sentence or selectively replace verbs. The overall set of prompts studied amounts to \totalQueries.

\subsection{Collection of the Results}

To complete our benchmark, we need to collect the code generated by Copilot for all the prompts we generated. To this end, we automatically generated a query package for each prompt in the dataset. The query package consists of a Java file with the prompt that has to be studied, followed by the signature of the method whose code has to be generated, and an empty method body implementation. For the cases derived from GitHub, we included in the file with the empty method the rest of the code of the class where the method occurs, so that Copilot could still exploit contextual information. 

To automatically collect the results generated by Copilot, since an API is not available, we implemented an automation tool with PyAutoGUI ({\small \url{www.pyautogui.readthedocs.io}}), which is a Python library for automating I/O operations. To implement a quick and reliable tool, we interacted with Copilot within Visual Studio Code mainly using shortcuts. In particular, our tool opens Visual Studio Code directly on the file with the empty method from the command line, it then searches that method,  moves the cursor over the target method, closes the search menu, enters into the empty body method, and invokes Copilot using shortcuts. Finally, it continuously takes screenshots until Copilot has produced a response, it then moves over the suggestion, accepts it, saves the files, and closes the editor.

The experimental material, including the full set of prompts and responses produced by Copilot, are available online at \urlRepo to facilitate further studies on this subject.

\subsection{Analysis of the Results}

To answer RQ1-3, we analyzed the code generated by Copilot according to multiple perspectives.

To answer RQ1, we checked if the generated code compiles and, in case it compiles, we determine if it also passes the execution of the available test cases. Although testing cannot guarantee the full correctness of the generated code, a code that passes a test suite that extensively exercises the code implemented by the developer is likely a quite useful piece of code.  
To determine if a prompt feature has a significant influence on the level of correctness of the code, we compute the contingency table and use the $\chi^2$ test to check significance.

To answer RQ2, we compute the cyclomatic complexity and the number of lines of code in the generated code using JaSoMe ({\small\url{https://github.com/rodhilton/jasome}}).  To determine if a prompt feature has a significant influence on the complexity and size of the generated code, we use the Wilcoxon-Mann-Whitney U test~\cite{Mann-Whitney} for prompt features with two categorical values and the Kruskal-Wallis H-test~\cite{Kruskal} for the prompt features with three or more categorical values.

To answer RQ3, we compare the generated code and the original code written by the developers according to both the normalized Levenshtein distance and the CodeBLEU metric~\cite{CodeBLEU}. The Levenshtein distance captures the percentage of syntactic changes necessary to obtain the desired code from the code generated by Copilot (the higher, the more diverse the two code snippets are). The CodeBLEU metric is a more sophisticated metric that captures the degree of syntactic and semantic similarity between the compared codes, considering both their abstract syntax tree and their data-flow (the higher, the more similar they are). We used the javalang library ({\small \url{https://pypi.org/project/javalang/})} to compute the normalized Levenshtein between tokens and the CodeXGLUE tool ({\small \url{https://github.com/microsoft/CodeXGLUE/tree/main}}) to compute the CodeBLEU metric.
To determine if the prompt features have a significant influence on the Levenshtein distance and on the CodeBLEU metric of the generated code, we use the Wilcoxon-Mann-Whitney U test~\cite{Mann-Whitney} for prompt features with two categorical values and the Kruskal-Wallis H-test~\cite{Kruskal} for the prompt features with three or more categorical values. 

For all the statistical tests, we consider a significance level of $0.05$, but we also mention when results are nearly significant, that is, they are significant at a significance level of  $0.1$. \section{Results} \label{sec:results}

In this section, we discuss the results obtained for the three research questions.

\subsection{RQ1 - Level of Correctness}

This research question investigates the correlation between the level of correctness of the code generated by Copilot and the content and style of the prompts, depending on the occurrences of the prompt features.

\begin{table}[ht]
\centering
\caption{Level of correctness of the generated code.}
\label{tab:correctness}
\resizebox{\columnwidth}{!}{
\begin{tabular}{lcccc}
\toprule
\textbf{Subject Cases} & \textbf{Tot. Prompts} & \textbf{No Compile} & \textbf{Fail Test} & \textbf{Pass All Tests}\\
\midrule
GitHub & 9,600& 6,246 (65\%) & 1,772 (18\%) & 1,582 (17\%)\\
LeetCode & 115,200 & 86,847 (75\%)& 14,434 (13\%)& 13,919 (12\%)\\
\midrule
\bf{Total} & 124,800 & 93,093 (75\%) & 16,206(13\%) & 15,501(12\%)\\
\bottomrule
\end{tabular}
}
\end{table}

\begin{figure}[ht]
 \begin{center}
   \includegraphics[width=1\columnwidth]{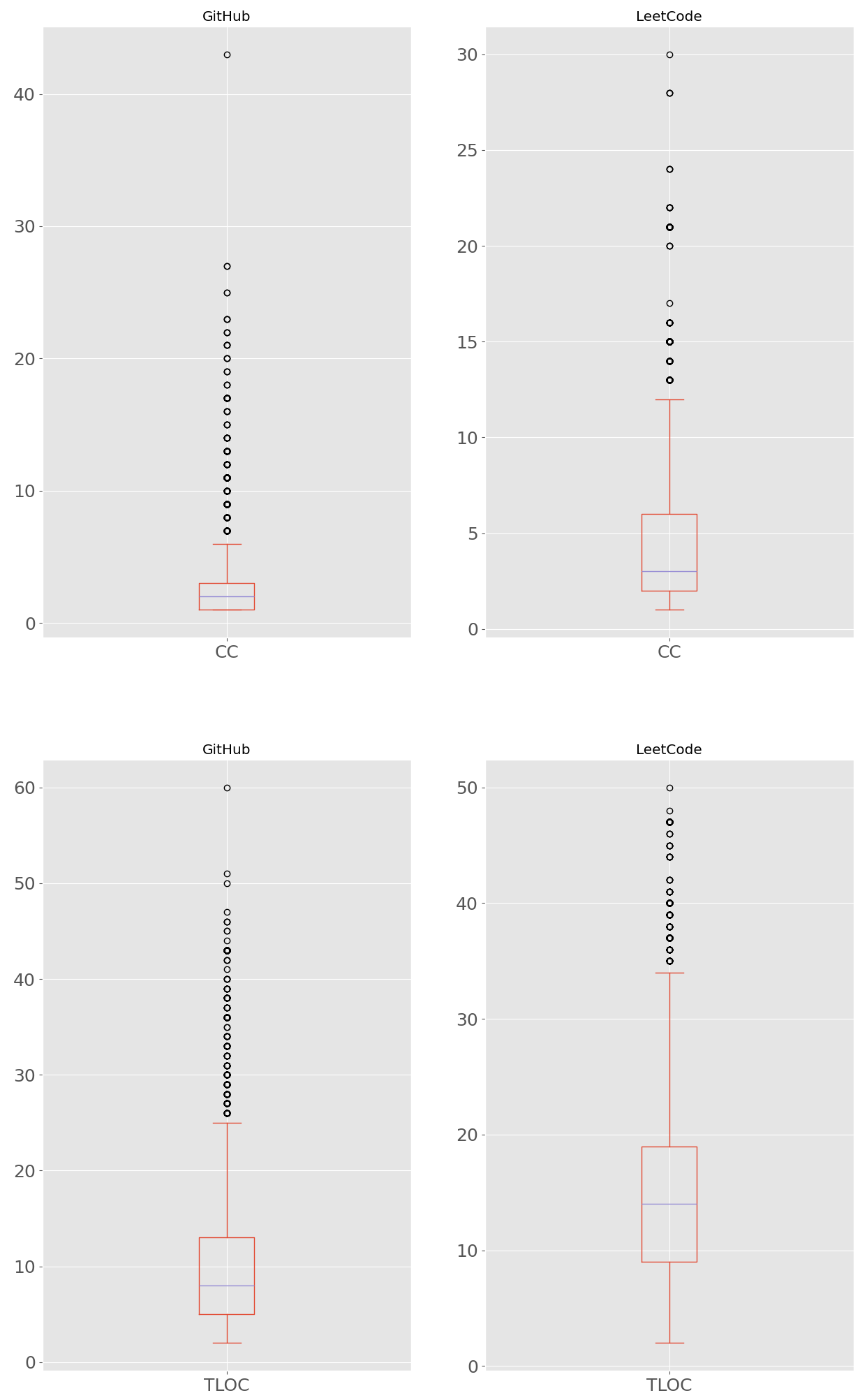}
\end{center}
   \caption{Complexity and size of methods in GitHub and LeetCode.}  \label{fig:CCLoc}
\end{figure}

Table~\ref{tab:correctness} reports the number and percentage of method bodies generated by Copilot that do not compile (Column \emph{No Compile}), that fail at least a test case (Column \emph{Fail Test}), and that pass all the available test cases (Column \emph{Pass All Tests}). The code generated for the GitHub methods compiled and passed test cases more frequently than the code generated for LeetCode methods: 35\% of method bodies compiled and 17\% of method bodies passed all the test cases for GitHub methods, while 25\% of method bodies compiled and 12\% passed all the test cases for LeetCode.  This is likely due to the different characteristics of the problems present in LeetCode, compared to the code present in actual GitHub projects. Figure~\ref{fig:CCLoc} shows the complexity and size of the GitHub and LeetCode methods in our benchmark. LeetCode methods tend to be more complex and longer than GitHub methods, likely explaining the difference in the results. GitHub methods capture the small-scale problems that developers face when writing code, while LeetCode methods capture some algorithmic problems that developers may have to face.  Overall, the percentage of cases that pass all the available test cases is quite limited (12\%), but consistent with previous results reported in studies about the effectiveness of Copilot. For instance, Mastropaolo et al.~\cite{Mastropaolo:RobustnessGenCode:ICSE:2023} obtained approximately 13\% method bodies that pass the test cases with Copilot using both the original prompts and those generated, either manually or automatically, through specialized tools.

\begin{table}[ht]
\centering
\caption{P-values of prompt features significantly influencing the level of correctness of the generated code.}
\label{tab:SigCorrectness}
\resizebox{\columnwidth}{!}{
\begin{tabular}{lccccc}
\toprule
\multirow{2}{*}{\textbf{Prompt Features}}& \multicolumn{2}{c}{\textbf{GitHub}} &  \phantom{ab} & \multicolumn{2}{c}{\textbf{LeetCode}} \\
\cmidrule{2-3} \cmidrule{5-6} 
& \textbf{Compile} & \textbf{Pass Tests} && \textbf{Compile} & \textbf{Pass Tests}\\

\midrule
Mood  & No & No && No& No\\
Mode  & No& No && No& No\\
Tense  & \cellcolor[gray]{0.8}0.1& \cellcolor[gray]{0.8}0.1 && No& No\\
\midrule
Param  & No& No&& No& No\\
B. Cases  & No& No && No& No\\
Summary  & - & - && No & \cellcolor[gray]{0.6}\textbf{0.04}\\
Examples  & - &-  && \cellcolor[gray]{0.6}$\mathbf{10^{-4}}$ & \cellcolor[gray]{0.6}$\mathbf{10^{-4}}$ \\
Context  & -& - && No& No\\

\bottomrule

\end{tabular}
}
\end{table}

\begin{figure}[ht]
 \begin{center}
   \includegraphics[width=\columnwidth]{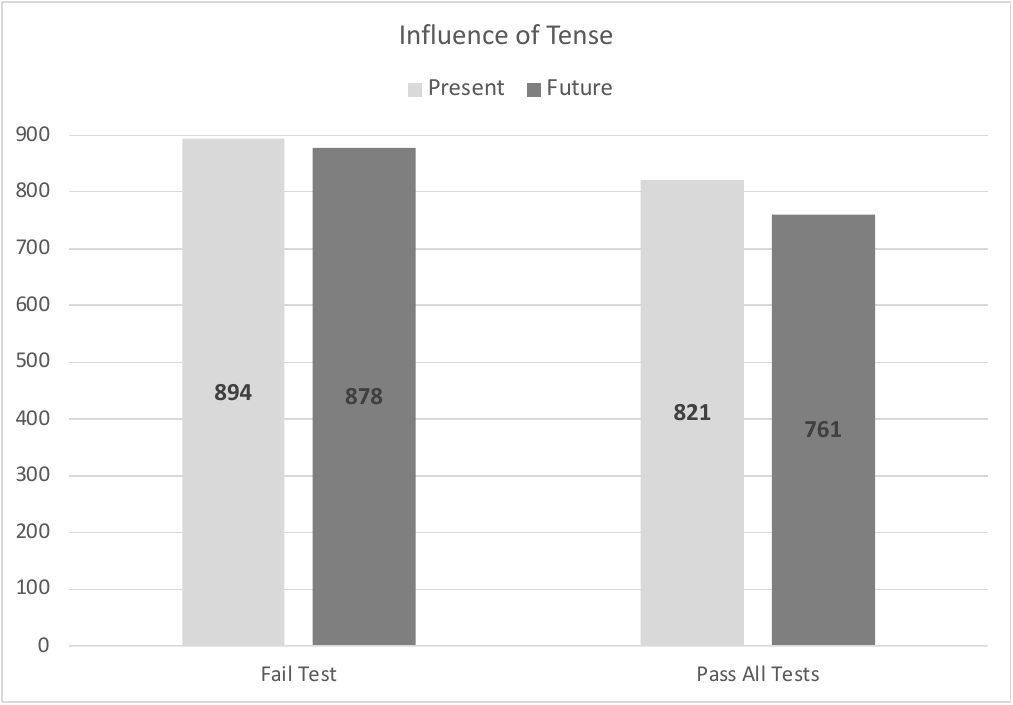}
\end{center}
   \caption{Influence of Tense on GitHub prompts.}  \label{fig:TenseFGitHub}
\end{figure}

In terms of the influence of the prompt features, Table~\ref{tab:SigCorrectness} lists all the considered features, and their significance on the generation of code that respectively compiles and passes all the available test cases, for both GitHub and LeetCode. We report \textit{No} when the prompt feature is not significant, we report the boldface p-value with a grey background when the feature is significant, we use a lighter background when the p-value is nearly significant, and \texttt{-} when the prompt feature does not apply. 

None of the five studied features have a statistically significant impact on the level of correctness of the results obtained for the prompts derived from GitHub. The only prompt feature with a mild influence on the correctness is the \emph{Tense} (p-value $0.1$). Figure~\ref{fig:TenseFGitHub} shows how \emph{Tense} impacts on the results. The present tense tends to be better interpreted by Copilot than the future tense, especially in terms of code that passes all the test cases. 

\begin{figure*}[!htb]
\minipage{0.32\textwidth}
  \includegraphics[width=\linewidth]{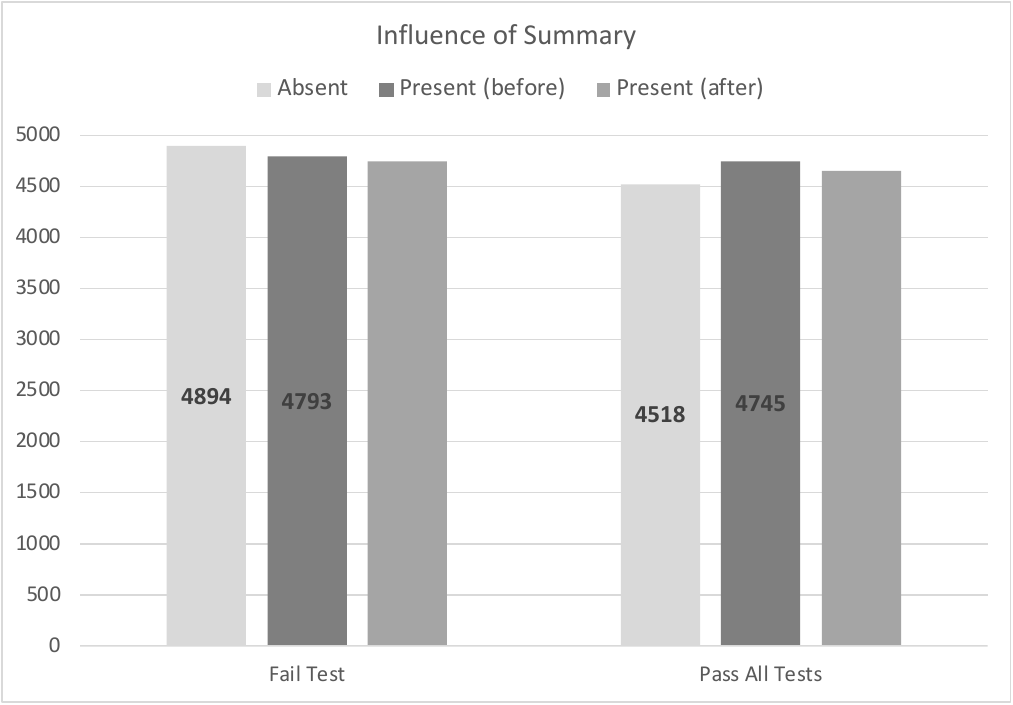}
  \caption{Influence of Summary on testing.}\label{fig:summary}
\endminipage\hfill
\minipage{0.32\textwidth}
  \includegraphics[width=\linewidth]{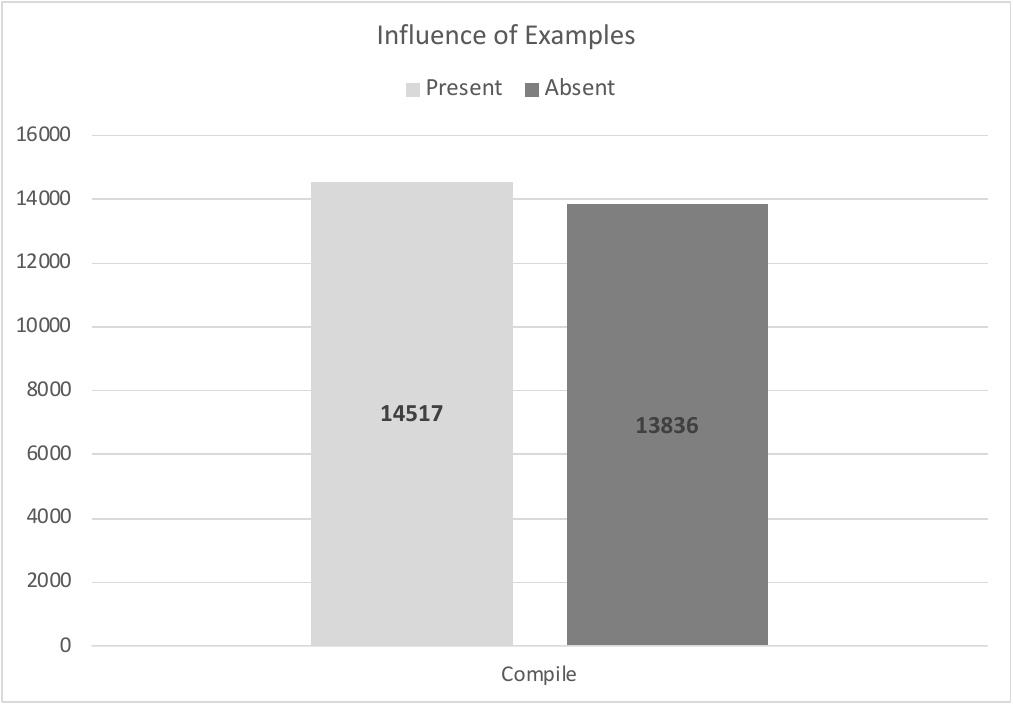}
  \caption{Influence of Examples on compilation.}\label{fig:examplesCompile}
\endminipage\hfill
\minipage{0.32\textwidth}\includegraphics[width=\linewidth]{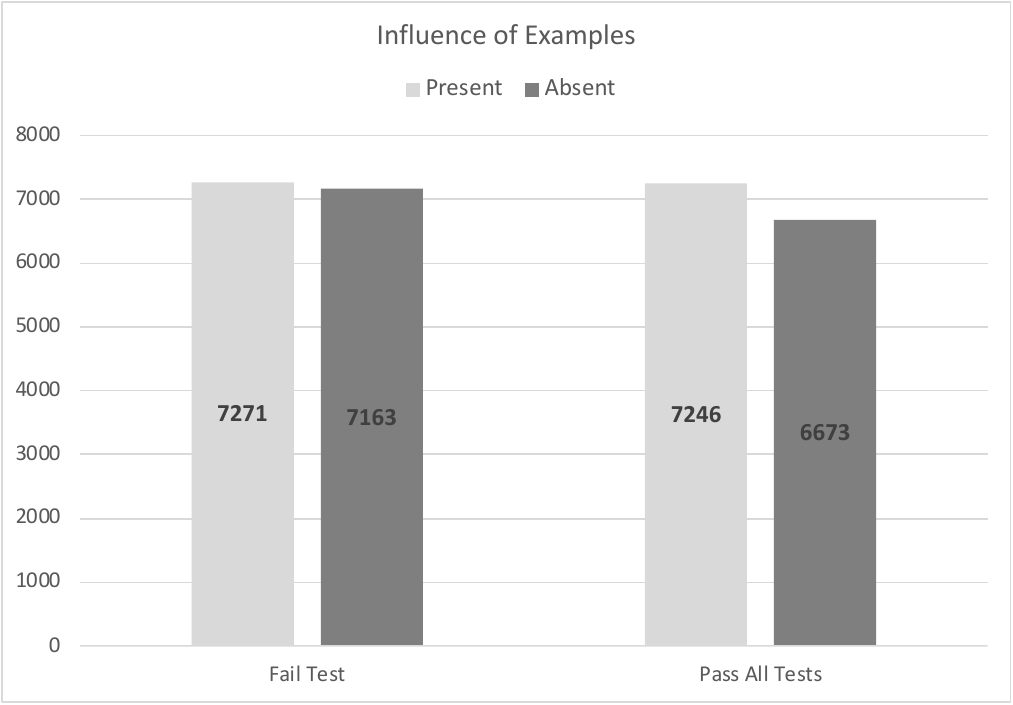}
  \caption{Influence of Examples on testing.}\label{fig:examplestest}
\endminipage
\end{figure*}

In the case of LeetCode, two prompt features have been statistically significant: the presence of a \emph{Summary}, for the purpose of obtaining code that passes all the available test cases only, and the presence of \emph{Examples}, for both the generation of code that compiles and passes all the tests. The \emph{Tense} in LeetCode is not even mildly significant. Probably, the \emph{Tense} has little impact when the description is richer, as the one derived from LeetCode, but it has a slightly more significant impact when the prompt is short. 

The presence of a \emph{Summary} of the purpose of the method has been significantly useful for generating code that passes the test cases, while it is not fundamental to simply obtain code that only compiles. Figure~\ref{fig:summary} shows the impact of \emph{Summary} that, especially when occurring before the rest of the description, increases the number of methods that pass all the available tests. 

The presence of \emph{Examples} resulted to be a key element for the success of the code generation task. Figures~\ref{fig:examplesCompile} and~\ref{fig:examplestest} show its impact on the generation of code that compiles and passes all the tests, respectively.  Indeed, examples provide important guidance for the generation of the right code for Copilot, especially to obtain code that passes the test cases. This is partially surprising since Copilot just interprets the text, not using the examples, for instance to run the generated code, and thus their strong impact on the results was not entirely expected.

In our experiments, the presence of \emph{Context} information about how methods are used, the specification of \emph{Boundary Cases}, the \emph{Reference to Parameters}, as well as the \emph{Mood} and the \emph{Sentence Mode}, have not revealed as introducing any significant difference on the level of correctness of the results. Based on our observations, they might tend to overcomplicate the prompt with information that cannot be fully exploited by the LLM.   
\smallskip

\begin{mdframed}
\textbf{Answer to RQ1} According to our results, effective prompts start with a \emph{summary} of the purpose of the method and include some \emph{examples} of its input/output behavior. Prompts should preferably use the \emph{present tense}. Providing additional contextual information does not necessarily produce significantly better results.
\end{mdframed}

\subsection{RQ2 - Complexity and Size}
This research question investigates the correlation between the complexity and size of the generated code, and the \emph{content} and \emph{style} of the considered prompts.

Table~\ref{tab:SigComplexSize} reports the prompt features that have a significant influence on code complexity and size. In the case of GitHub, the \emph{Sentence Mode} has been reported as significantly influencing complexity, although generating a mild impact. In fact, the mean (and standard deviation) of the complexity when using the active mode is $2.81(\pm 2.5)$, while when using the passive mode is $2.72(\pm 2.35)$. It thus seems that using the passive form may mildly influence code complexity, when the prompt is short (e.g., the GitHub prompt) and does not include additional information such as \emph{Examples}, \emph{Summary of the Method}, and \emph{Boundary Cases}, as opposed to LeetCode's prompts. Due to difficulty in explaining observations with LLM, it is hard to identify the reason for such a mild dependency.

\begin{table}[ht]
\centering
\caption{P-values of prompt features significantly influencing the complexity and size of the generated code.}
\label{tab:SigComplexSize}
\resizebox{\columnwidth}{!}{
\begin{tabular}{lccccc}
\toprule
\multirow{2}{*}{\textbf{Prompt Features}}& \multicolumn{2}{c}{\textbf{GitHub}}  &\phantom{ab} & \multicolumn{2}{c}{\textbf{LeetCode}} \\
\cmidrule{2-3} \cmidrule{5-6}
& \textbf{Complexity (CC)} & \textbf{Size (loc)} && \textbf{Complexity (CC)} & \textbf{Size (loc)}\\
\midrule
Mood  & No & No && No& No\\
Mode  & \cellcolor[gray]{0.6}\textbf{0.02}& No && No& No\\
Tense  & No& No&& No& No\\
\midrule
Param  & No& No&& No& No\\
B. Cases  & \cellcolor[gray]{0.6}$\mathbf{10^{-6}}$& \cellcolor[gray]{0.6}$\mathbf{10^{-1}}$ && No& No\\
Summary  & - & - && \cellcolor[gray]{0.8}0.09 & No\\
Examples  & - &-  && \cellcolor[gray]{0.6}$\mathbf{10^{-11}}$ & \cellcolor[gray]{0.6}$\mathbf{10^{-23}}$ \\
Context  & -& - && No& No\\

\bottomrule

\end{tabular}
}
\end{table}

The specification of the \emph{Boundary Cases} had a significant influence on the complexity and size of the code generated for the GitHub cases. 

\begin{figure}[ht]
 \begin{center}
\includegraphics[width=1\linewidth]{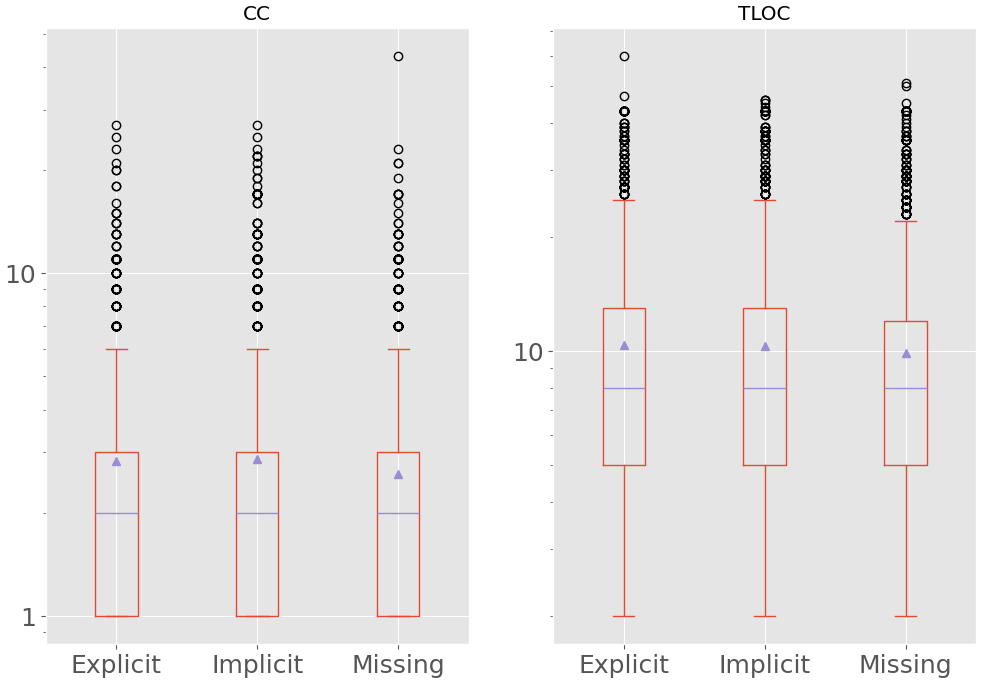}
    \end{center}
   \caption{Influence of Boundary Cases on complexity and size.}  \label{fig:BoundOnCCSize}
\end{figure}

Figure~\ref{fig:BoundOnCCSize} shows how the presence of the \emph{Boundary cases} induces the generation of longer and slightly more complicated code. In fact, when a boundary case is explicit, Copilot often generates code that explicitly incorporates \texttt{if} statements to handle the boundary cases specified in the prompt. For example, one of the parameters of the method \texttt{defineLigand}, which is one of the methods that we selected from GitHub, is an object of type \texttt{IAtomContainer} named \texttt{container}. The implementation is supposed to check the parameter for \texttt{null} values, and Copilot adds this check in the code only when the boundary case is specified. The presence of the check increases the level of correctness, but also the length and the complexity of the resulting code. While Copilot often outputs additional checks when boundary cases are included in the prompt, thus increasing the size and complexity of the code, these extra checks are not sufficient to significantly increase the level of correctness of the code in our study, since they are sometimes unnecessary or implied by the already generated code (see the results for RQ1). 

When the prompt is particularly rich, for instance it already includes summaries and examples, the inclusion of boundary cases does not have an impact on the complexity and size of the code, as reported by the lack of significance of the boundary cases on the size and complexity of the code generated for LeetCode's methods. This is likely to happen because examples and boundary cases report partially overlapping information.

\begin{figure}[ht]
 \begin{center}
\includegraphics[width=\linewidth]{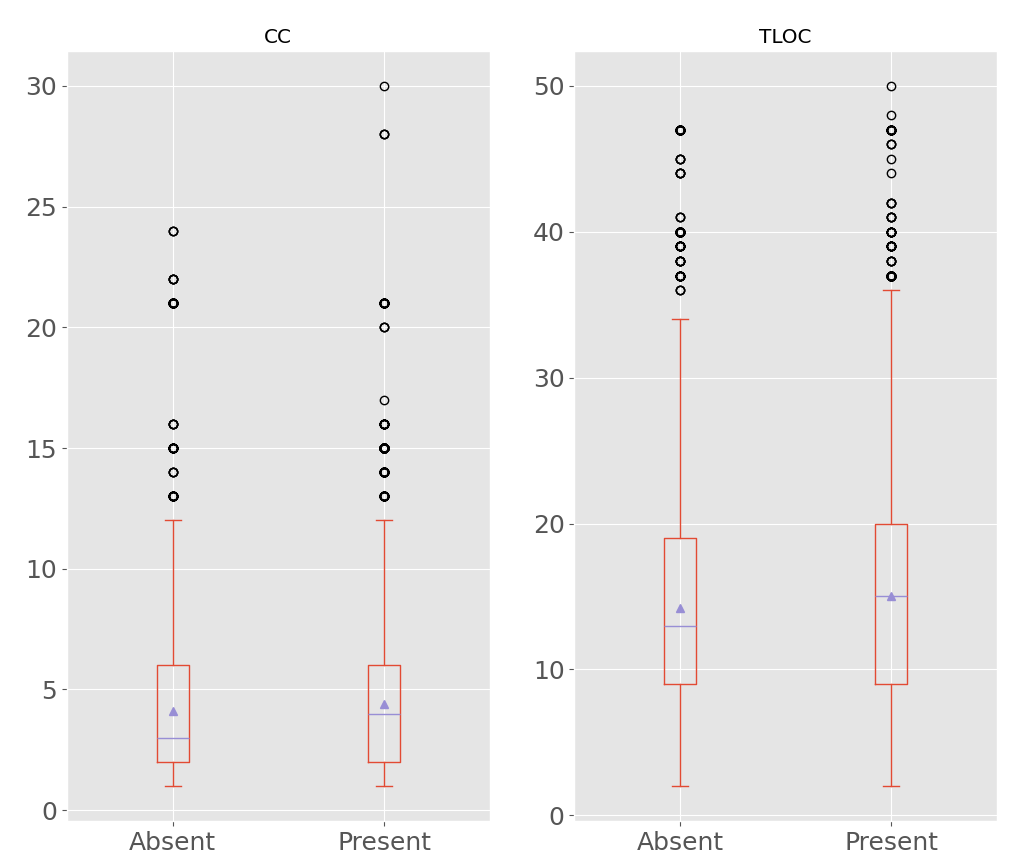}
    \end{center}
   \caption{Influence of Examples on 	complexity and size.}  \label{fig:ExampleonCCSize}
\end{figure}

\begin{table}[ht]
	\centering
	\caption{Mean size and complexity of the generated code.}
	\label{tab:complexsize}
	\resizebox{\columnwidth}{!}{
		\begin{tabular}{llcccc}
			\toprule
			\textbf{Subject} & \textbf{Metric} & \textbf{Not Compile} & \textbf{Compile} & \textbf{Fail Test} & \textbf{Pass All Tests}\\
			\midrule
			\multirow{ 2}{*}{GitHub} & CC & 2.68& 2.87 & 3.06 & 2.65\\
			 & Locs & 9.48 & 11.15& 11.42& 10.84\\
			\midrule
			\multirow{ 2}{*}{LeetCode} & CC & 1.88& 4.85 & 5.05 & 4.85\\
			 & Locs & 6.28 & 16.70& 17.09& 16.31\\
			 \bottomrule
		\end{tabular}
	}
\end{table}

The \emph{Summary} and especially the \emph{Examples} had a significant influence on code complexity and size of the generated code. Figure~\ref{fig:ExampleonCCSize} quantifies the impact of these prompt features. This result was expected since these prompt features correlate with the correctness of the code and the largely wrong code (i.e., code that does not even compile) is often over-simplistic. Table~\ref{tab:complexsize} reports the complexity (row \emph{CC}) and size (row \emph{Locs}) for the methods generated for both GitHub and LeetCode cases, distinguishing between the methods that do/do not compile and, among the ones that compile, the ones that fail/do not fail at least a test case. These results show how the shortest code usually does not even compile. In fact, the code that does not compile has the lowest mean complexity and the smallest mean size for both GitHub and LeetCode. On the other hand, it is interesting to see how, among the code that compiles, the most complex and largest one usually fails at least a test case. In fact, the highest mean complexity and largest mean size are reported for the code that fails at least a test for both GitHub and LeetCode.    

\begin{table}[ht]
\centering
\caption{Significant relationships between size and complexity of the code to be generated and the level of correctness of the solution.}
\label{tab:complexsizeOriginalCode}
\resizebox{\columnwidth}{!}{
\begin{tabular}{llcc}
\toprule
\multirow{ 2}{*}{\textbf{Subject}} & \multirow{ 2}{*}{\textbf{Metric}} & \textbf{Generated Code} & \textbf{Generated Code} \\
 & & \textbf{Compiles} & \textbf{Passes All the Tests} \\
\midrule
\multirow{ 2}{*}{GitHub} & CC of the original code & No & \cellcolor[gray]{0.6}{$\mathbf{10^{-8}}$}  \\
 & Locs in the original code & \cellcolor[gray]{0.6}{$\mathbf{10^{-30}}$} & \cellcolor[gray]{0.6}{$\mathbf{10^{-32}}$}\\
\midrule
\multirow{ 2}{*}{LeetCode} & CC of the original code& \cellcolor[gray]{0.6}{$\mathbf{10^{-6}}$} & \cellcolor[gray]{0.6}{$\mathbf{0.001}$}  \\
 & Locs in the original code& \cellcolor[gray]{0.6}{$\mathbf{10^{-8}}$} & \cellcolor[gray]{0.6}{$\mathbf{10^{-15}}$}\\
\bottomrule
\end{tabular}
}
\end{table}

\begin{figure}[ht]
 \begin{center}
      \includegraphics[width=\linewidth]{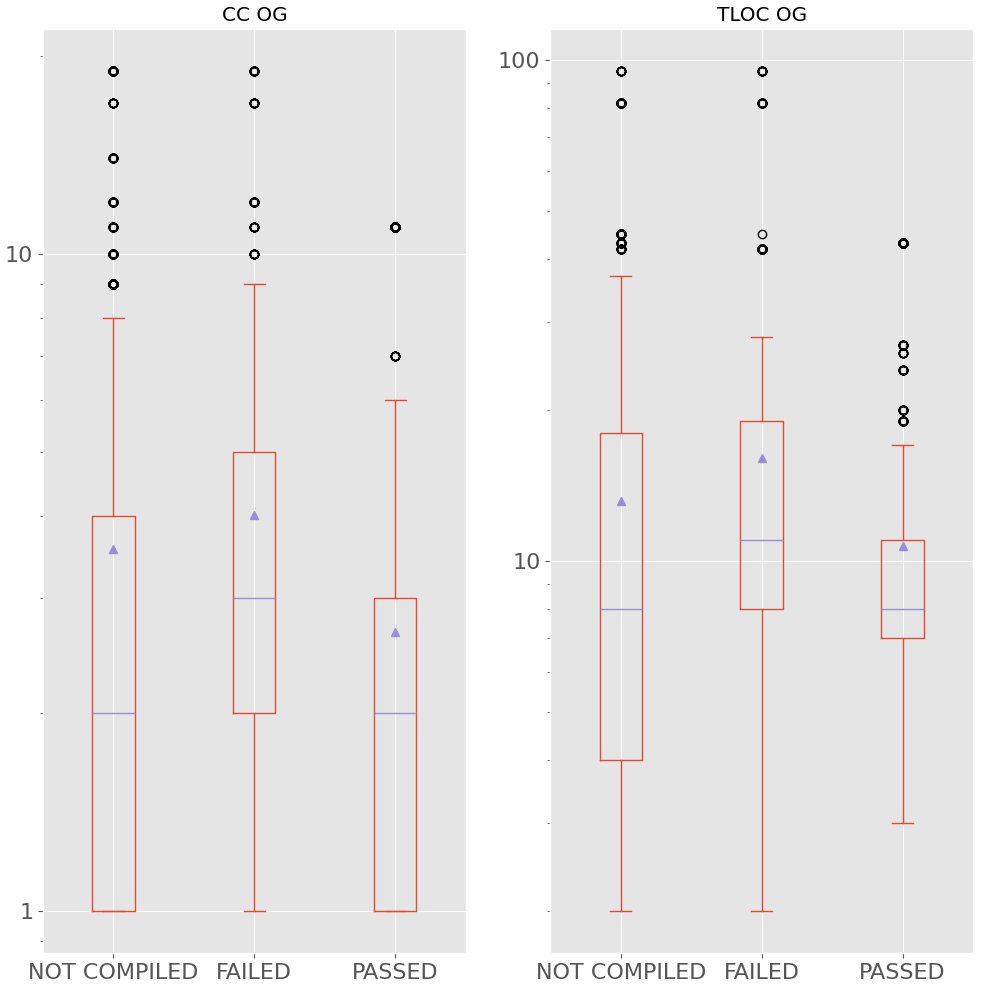}
    \end{center}
   \caption{Complexity and size of the developers' code for the GitHub methods whose code generated by Copilot does not compile, fails at least a test case, or passes all the tests.}  \label{fig:CCSizeOrigGitHub}
\end{figure}

\begin{figure}[ht]
 \begin{center}
      \includegraphics[width=\linewidth]{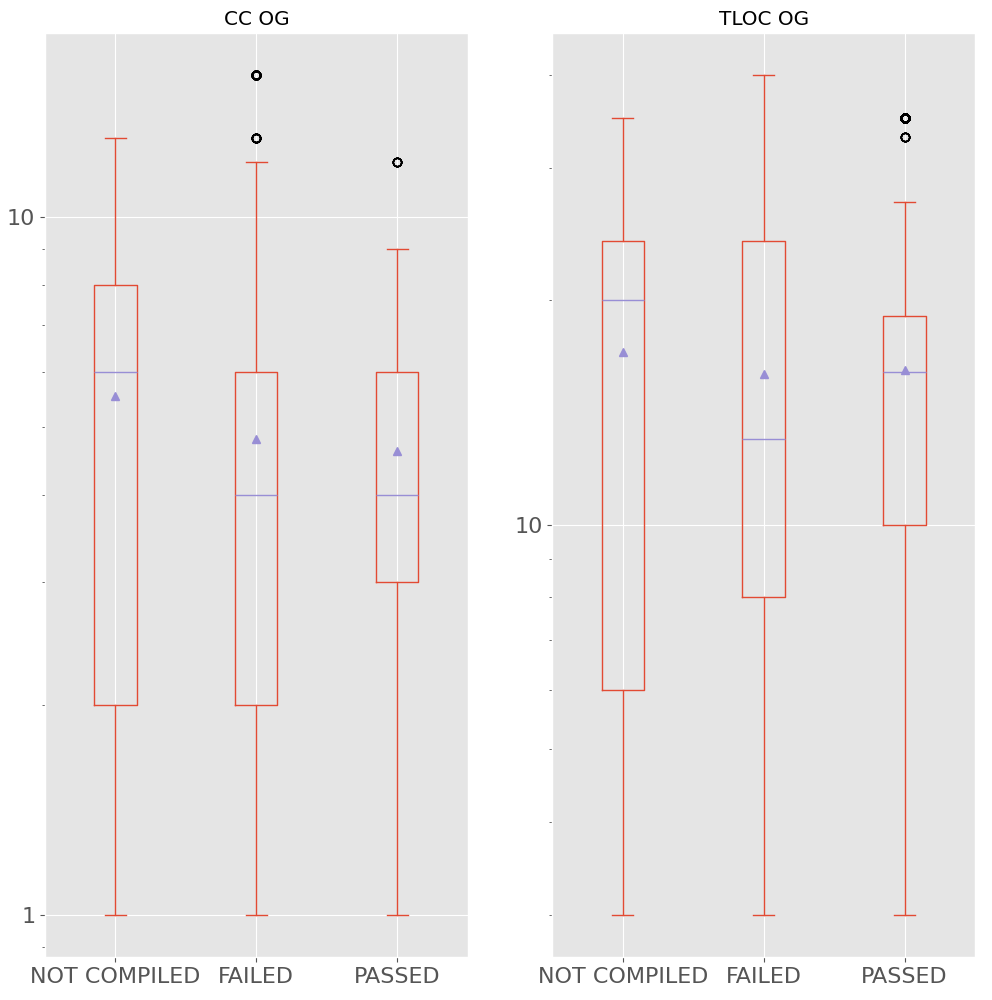}
    \end{center}
   \caption{Complexity and size of the developers' code for the LeetCode methods whose code generated by Copilot does not compile, fails at least a test case or passes all the tests.}  \label{fig:CCSizeOrigLeetCode}
\end{figure}

We also investigated if a similar relationship holds with the code that has to be generated, that is, the complexity and length of the original code extracted from GitHub and LeetCode. Table~\ref{tab:complexsizeOriginalCode} shows the significant correlations between complexity and size metrics and the resulting level of correctness of the code, distinguishing between achieving code that compiles and code that passes all the test cases. We can notice that there is a strong correlation between all these factors (all the correlations are significant with the exception of code complexity correlated to the generation of code that compiles in GitHub).

Figures~\ref{fig:CCSizeOrigGitHub} and~\ref{fig:CCSizeOrigLeetCode} show the results for GitHub and LeetCode, respectively. Interestingly, they show opposite trends for GitHub and LeetCode, concerning the code that fails to compile. Although the population is spread, most of the code that fails to compile originates from the attempt to generate the body of a method that is rather simple and short in GitHub. While most of the code that fails to compile originates from the attempt to generate the body of methods that are rather complex and long in LeetCode. This can be explained by the nature of the cases. The complexity of GitHub is more on the use of appropriate APIs and less on the algorithmic aspects, making the complexity and length of a method a less relevant factor. While the length and complexity are a more important factor for the algorithmic code present in LeetCode.  

Once code that compiles is generated, Copilot succeeds mostly with the simpler and shorter methods in GitHub, while it is not necessarily the case in LeetCode.

\smallskip

\begin{mdframed}
\textbf{Answer to RQ2} The complexity and length of the generated code are influenced by the \emph{Sentence Mode} of the prompt, the presence of \emph{Boundary Cases}, the presence of the \emph{Summary} and the \emph{Examples}.  While the examples demonstrated beneficial for code correctness, it was not the same for the boundary cases. The mode had also a mild impact on the complexity of the generated code, with the passive form producing slightly simpler code.
\end{mdframed}

\subsection{RQ3 - Similarity to the Intended Code}

This research question investigates how close the generated code, even if incorrect, is to the original code according to the Levenshtein and the CodeBLEU metrics.
Table~\ref{tab:SigLevBleu} reports the significant prompt features for these two metrics.

\begin{table}[ht]
\centering
\caption{P-values of prompt features significantly influencing the normalized Levenshtein distance and the CodeBLEU of the generated code.}
\label{tab:SigLevBleu}
\resizebox{\columnwidth}{!}{
\begin{tabular}{lcclcc}
\toprule
\multirow{2}{*}{\textbf{Prompt Features}} & \multicolumn{2}{c}{\textbf{GitHub}}  & \phantom{ab}&  \multicolumn{2}{c}{\textbf{LeetCode}} \\
\cmidrule{2-3} \cmidrule{5-6} 
& \textbf{Levenshtein} & \textbf{CodeBLEU} && \textbf{Levenshtein} & \textbf{CodeBLEU}\\
\midrule
Mood  & No & No && No& No\\
Mode  &No & No && \cellcolor[gray]{0.6}$\mathbf{0.05}$ & \cellcolor[gray]{0.6}$\mathbf{0.01}$\\
Tense  & No& No&& No& No\\
\midrule
Param  & No& No&& No& No\\
B. Cases  & \cellcolor[gray]{0.6}$\mathbf{0.007}$ & \cellcolor[gray]{0.8}$\mathbf{0.054}$ && No& No\\
Summary  & - & - && No & No\\
Examples  & - &-  && No & No\\
Context  & -& - && No& No\\

\bottomrule
\end{tabular}
}
\end{table}

\begin{figure}[h]
 \begin{center}
\includegraphics[width=1\linewidth]{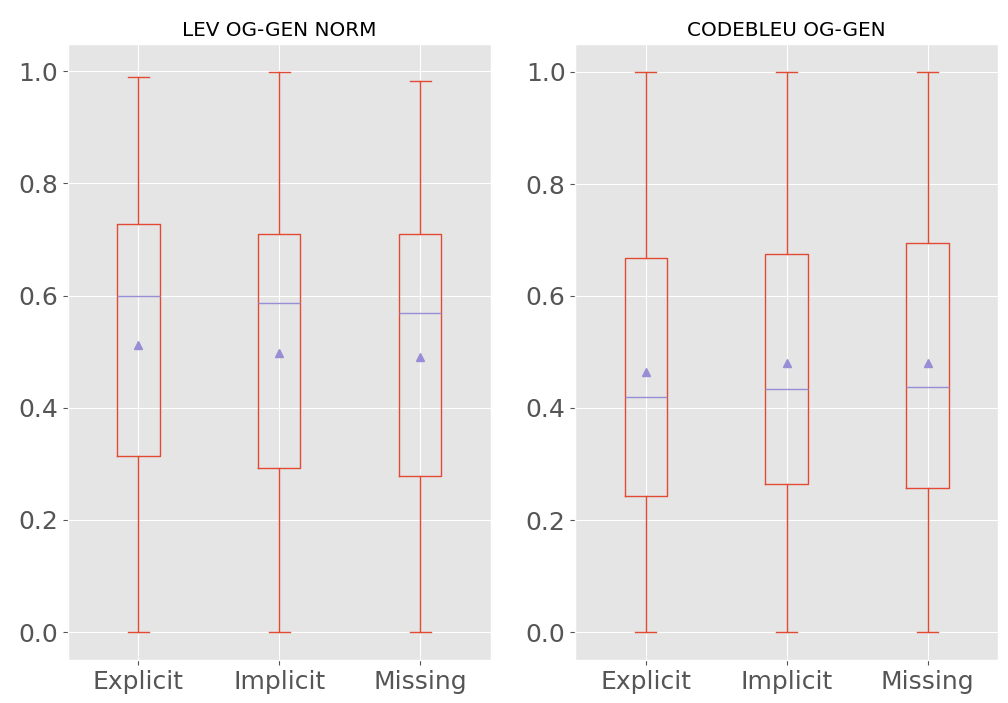}
    \end{center}
   \caption{Influence of Boundary Cases on Levenshtein distance and CodeBLEU.}  \label{fig:boundaryLevBlueC}
\end{figure}

Results show that the \emph{Sentence Mode} has an influence on the Levenshtein distance and the CodeBLEU distance for the LeetCode cases. In particular, when the active mode is used, the mean Levenshtein distance is $0.601$ while it is $0.599$ with the passive mode. Similarly, the CodeBLEU is $0.412$ with the active mode and $0.415$ with the passive one. The active mode generates code that is slightly more diverse from the original code according to both metrics (higher Levenshtein distance and lower CodeBLEU). However, the impact of this phenomenon is marginal.

The presence of \emph{Boundary cases} also had a significant impact on the similarity of the resulting code compared to the original code. Figure~\ref{fig:boundaryLevBlueC} shows the influence of boundary cases. They cause the code to differ slightly more from the original code (higher Levenshtein distance and lower CodeBLEU). This suggests that boundary cases might induce the generation of conditions that should not be present in the code, increasing the level of diversity between the generated code and the original one. Considering that boundary cases do not have a positive effect on correctness (RQ1) and tend to generate longer code (RQ2), this result confirms that it might be better not to include them in the prompt, representing a possible overload of information for the LLM.

Note that improving the similarity between the generated code and the original code is important also when the generated code is not fully correct, since developers will likely have to implement fewer changes to obtain satisfactory code.

\smallskip

\begin{mdframed}
\textbf{Answer to RQ3} Although none of the prompt features had a large influence on the level of similarity between the code generated by Copilot and the original code, some prompt features may have a mild influence. In particular, using the \emph{passive mode} and not stating \emph{Boundary Cases} explicitly may result in code with higher similarity to the original one.
\end{mdframed}

\subsection{Threats to Validity}
As with every study, also our empirical study is affected by multiple threats to validity that we identified and addressed as follows. 

The main internal threat to validity is about the actual influence that prompt features had on the results. We addressed this threat in two ways. First, we studied prompt features systematically, considering every possible combination. Second, we investigated their impact statistically, to focus on the relevant phenomena. Although we cannot claim the presence of any strong cause-effect relationship between inputs and outputs, due to the low explainability of large language models, the results reported in this paper provide an initial guidance towards the challenge of designing good prompts. 

Another threat is how we measure the quality of the generated code. We relied on automated methods to assess the quality of the generated code due to the scale of the study (124,800 methods generated with Copilot). In particular, we relied on the capability to compile the code, and the execution of test cases, to capture different levels of correctness of the code. Indeed, we never claim that a method that passes all the available test cases is a fully correct method, but yet test execution provides useful information about the quality of the code. This limitation is shared with several other studies in the area~\cite{Nguyen:EmpiricalEvaluationCopilot:MSR:2022,Chen:EvaluatingLLMs:arXiv:2023,Li:CCTEST:ICSE:2023,Yetistiren:AssessingCopilot:PROMISE:2022}. 

To not limit the analysis to test execution, we also consider a combination of syntactic (Levenshtein distance) and semantic (CodeBLEU) metrics that measure the differences between the two methods, disclosing information about the level of closeness of the generated code to the original developers' code.

The main external threat to validity concerns the representativeness and generalizability of the findings. Our study targets a common, although specific, use case, that is, the generation of the body of a method. Results cannot be generalized beyond this use case. Moreover, we focus on the design of a good prompt with the aim of obtaining the right code with a single request. Engineering conversations with large language models is a different problem. Again, although our findings might be useful also in the context of a conversation, the study explicitly focuses on individual and independent interactions (e.g., useful to decide how to start the conversation). Finally, we covered both the case of Java methods extracted from GitHub, representing small-scale problems that developers face daily, and the case of Java methods extracted from LeetCode, representing algorithmic problems that developers may also have to face. We do not know if our findings can be generalized beyond these two cases, which however already cover a reasonable spectrum of the interesting coding problems faced by developers.

\subsection{Prompt Advices}

Based on our results, there are some advices that could be exploited to design an initial prompt to request for some code. Although the study focuses on individual features, and not on the impact of multiple interacting features, we can conjecture that it is useful to include the following three sections in the prompt:  

\texttt{<Summary> <Description (Present Tense)><Examples>}

The prompt should start with a short summary of the purpose of the method. It follows a description of the behavior of the method, preferably written using the present tense (with either imperative or indicative mood). The usage of the passive mode may sometimes reduce the length of the generated code. Finally, the examples should include a few input-output pairs that demonstrate how the code should behave in some specific cases. 

The following listing shows an example prompt extracted from our dataset that matches this structure. We marked the three main sections with \verb|<summary>|, \verb|<description>|, and \verb|<examples>|.

\begin{Verbatim}[frame=single,fontsize=\footnotesize]
<summary>
Return true if any match is found.
</summary>

<description (present tense)>
Given an input string (s) and a pattern (p), the function  
implements wildcard pattern  matching with support for '?' and '*' 
where:
'?' Matches any single character.
'*' Matches any sequence of characters (including the empty 
    sequence).
The matching should cover the entire input string (not partial).
</description>

<examples>
Example 1:
Input: s = "aa", p = "a"
Output: false
Explanation: "a" does not match the entire string "aa".

Example 2:
Input: s = "aa", p = "*"
Output: true
Explanation: '*' matches any sequence.

Example 3:
Input: s = "cb", p = "?a"
Output: false
Explanation: '?' matches 'c', but the second letter is 'a', which 
does not match 'b'.
</examples>
\end{Verbatim}
 \section{Related Work}
\label{sec:related}

The use of generative models of artificial intelligence, such as language models and code generators, has gained considerable attention in recent years, and is changing the way developers create software. Generative AI solutions, and in particular those built on LLMs have demonstrated remarkable success in understanding and generating code for various programming languages~\cite{Li:Alphacode:Science:2022, Nijkamp:CodeGen:ICLR:2023, Fried:InCoder:ICLR:2023}. Some other recent studies investigated how developers interact with these tools during project development~\cite{Barke:CoPilotGroundTheory:ACMPL:2023}, and how usable these tools are~\cite{Barke:CoPilotGroundTheory:ACMPL:2023}. However, the effectiveness of these models largely depends on the quality and informativeness of the provided prompts~\cite{Denny:ConversingCoPilot:TSCSE:2023}. There is a growing interest in understanding the influence of prompt features on the quality of the generated code.

Prompts are essential for guiding language models towards specific tasks and outputs, without the need for retraining \cite{Rodriguez:Prompts:REW:2023,Liu:PromptImages:CHI:2022,Lo:ArtPrompting:IRSQ:2023}. It can be difficult to understand a model's true capabilities and distinguish between infeasible tasks and those where the model simply misunderstood the prompt. A failed task may indicate a poorly designed prompt, rather than the model's inability to perform the task~\cite{Brown:FewShotLearners:NIPS:2020}. For instance, previous research has explored various aspects of prompt engineering. Brown et al.~\cite{Brown:FewShotLearners:NIPS:2020} observed that enriching prompts with practical examples of the desired task enhances the capabilities of these models, with more examples correlating positively with better output quality. This is consistent with our findings in the context of LLMs used to generate the body of methods. On the other hand, Reynolds and McDonell~\cite{Reynolds:PromptProrammin:CHI:2021} studied example-free approaches, focusing on prompt engineering to improve results, outperforming, in some cases, poorly structured prompts with examples. They utilized techniques like task specification, encompassing various methods to describe the same goal, both directly and through proxies, such as analogies and synonyms for common concepts. They also explored methods to restrict unwanted outputs. Other studies~\cite{White:PromptPatternCatalog:arXiv:2023, White:PromptPatterns:arXiv:2023} have introduced and categorized prompt patterns, similar to software design patterns, with the goal of establishing a context for models like ChatGPT and directing them toward specific desired outcomes. While these patterns were initially developed with software development in mind, their utility extends to a wide range of model applications. Moreover, some of these patterns rely on the conversational aspect of the model, which may not align with Copilot's non-conversational nature. Building upon this prior research, our study uniquely focuses on the impact of \numFeatures prompt features on the quality of  the code produced by Copilot.

In recent years, various studies have explored the capabilities and limitations of Copilot and its underlying model. Some initial evaluations focused on the classic approach of testing the correctness of the solutions provided~\cite{Nguyen:EmpiricalEvaluationCopilot:MSR:2022, Chen:EvaluatingLLMs:arXiv:2023, Austin:ProgramSynthesisLLMs:arXiv:2023, Corso:EmpiricalAssessment:ICPC:2024}. These assessments did not employ prompt engineering techniques, but observed that Copilot was capable of generating correct code for relatively simple tasks even without detailed implementation guidance. An alternative perspective was explored in a study that examined metamorphic testing of Copilot, focusing on prompt modifications~\cite{Li:CCTEST:ICSE:2023}. This research investigated the effect of altering prompts, with a unique emphasis on code fragments as prompts rather than natural language descriptions. The study conducted operations involving semantically equivalent mutations, often resulting in distinct code outputs. 

Finally, a recent empirical study analyzed the impact of changes made to prompts expressed in natural language through both manual and automated paraphrasing, leveraging tools such as PEGASUS~\cite{Zhang:Pegasus:ICML:2020} and translation pivoting~\cite{Mastropaolo:RobustnessGenCode:ICSE:2023}. The authors created a dataset of Java methods collected from established projects on GitHub. They maintained the original code while automatically generating code for these methods using paraphrases of the original descriptions as prompts for Copilot. The study revealed significant diversification in terms of code correctness, similarity to the original code, and complexity of the code produced, all driven by changes in the prompt. Denny et al.~\cite{Denny:ConversingCoPilot:TSCSE:2023} conducted research to assess the impact of prompt modifications when tackling Python programming challenges. Their findings suggest that prompt engineering has proven beneficial in enhancing the accuracy of the resulting code across a majority of cases. White et al.~ \cite{White:PromptPatterns:arXiv:2023} introduced various prompt patterns that programmers can utilize in different scenarios, although they did not provide data regarding the efficacy of these patterns. Additionally, Ren et al.~\cite{Ren:CodeGen:ASE:2023} focused on adjusting prompts to facilitate the generation of proper exception handling code. 

Nevertheless, to the best of our knowledge, prompt engineering in code generation tasks has not been studied systematically. Our work provides initial insights about prompt engineering for code generation, considering Copilot as generative AI solution.

 \section{Conclusions} \label{sec:conclusions}

Generative AI techniques represent sophisticated services that developers can exploit to increase their efficiency and effectiveness in several tasks, including code development. In this context, Copilot is one of the most used tools based on generative AI that can assist developers while they are coding from within the integrated development environment. A key capability of Copilot is the possibility to generate the body of a method starting from a prompt, that is, the request written by the developer as the method's comment. 

To obtain a good response from generative AI tools, it is well-known that crafting an appropriate request is mandatory. This paper systematically studies this aspect by empirically investigating how eight prompt features may impact the effectiveness of the prompts submitted to Copilot. The reported study involves 200 Java methods and 124,800 prompts, whose response is assessed in terms of their correctness, complexity, size, and similarity to the intended code. 

The results show how only some prompt features influence the results. In particular, we report how a good prompt should include a summary of the semantics of the method, a description of its behavior, possibly using the presence tense, and some examples of input-output pairs that exemplify the behavior of the method. On the contrary, other prompt features, such as the description of boundary cases, the reference to the parameters, and contextual information have little influence on the results. We ended up recommending a structure for the prompt.

Our results represent a starting point for a deeper investigation of this subject. In fact, our future work includes widening the study to consider conversations between the developers and the generative AI tools, and not only individual interactions. Further, we plan to extend our analysis to other tools to investigate to what extent prompt engineering, and in particular results about prompt features, generalize across generative AI solutions.   
\begin{acks}
This work has been partially supported by the Engineered MachinE Learning-intensive IoT systems (EMELIOT) national research project (PRIN 2020 program Contract 2020W3A5FY) and the MUR under the grant "Dipartimenti di Eccellenza 2023-2027" of the Department of Informatics, Systems and Communication of the University of Milano-Bicocca, Italy.
\end{acks}

\balance
\bibliographystyle{ACM-Reference-Format}

\end{document}